\documentclass{article}  
\usepackage{lajolla2006}
\usepackage{graphicx}
\frompage{000} \topage{000}                                              
\def\rightmark{Long-Range Mult. Corr. in Au+Au Collisions at $\sqrt{s_{NN}}$ = 200 GeV}
\def\leftmark{T. Tarnowsky (for the STAR Collaboration)}
 
\title{Long-Range Multiplicity Correlations in Au+Au at $\sqrt{s_{NN}}$ = 200 GeV} 
\authors{
{Terence J. Tarnowsky $^1$ (for the STAR Collaboration)
}\\[2.812mm]
{\normalsize
\hspace*{-8pt}$^1$ Purdue University, \\ 
525 Northwestern Avenue, West Lafayette, IN 47907, USA\\[0.2ex] 
}}
 
\abstract{We present data on long-range multiplicity correlations in ultra-relativistic heavy ion collisions at the top RHIC energy ($\sqrt{s_{NN}}$ = 200 GeV) from the STAR experiment. The data shows a long-range multiplicity correlation extending across a gap of 1.6 units in pseudorapidity. The data is over predicted by a multiparticle production model with hadronization of independent strings, or fusion of two color strings. This can be interpreted in terms of additional dynamical reduction in the number of particle sources.}

\keyword{STAR, RHIC, correlation, string fusion, forward-backward} 
\PACS{25.75.Dw, 25.75.Gz, 25.75.-q}
 
\begin{document}
 
\maketitle
\setcounter{page}{1}

\section{Introduction}\label{intro}

The measurement of particle correlations has been suggested as a method to search for the existence of a phase transition in ultra-relativistic heavy ion collisions \cite{bib1,bib2,bib3}. If the quark-gluon plasma (QGP) is formed in these collisions, the existence or absence of particle correlations could lead to a determination of the presence of partonic degrees of freedom. Multiparticle production at low $p_{T}$ in 200 GeV is well described by string hadronization models, such as the Parton String Model (PSM), a Monte Carlo realization of the Dual Parton Model (DPM), or Quark-Gluon String Model (QGSM), which includes fusion of overlapping color strings \cite{bib4,bib5,bib6}. The color strings are formed between the projectile and target nuclei. These strings cover a small area in the transverse collision plane that contains the color field created by colliding partons. The number of color strings is directly related to the atomic number or energy of the colliding nuclei increases, so does the number of color strings. The strings will then begin to overlap in the transverse plane, forming clusters. The PSM (an evolution of the String Fusion Model \cite{bib18}), incorporates the fusion of two strings produced from soft collisions. 

The transverse area of the strings goes as $\displaystyle\frac{1}{\left<p{_T}^{2}\right>}$, so there is a low probability of fusing hard strings (strings with large mean transverse momentum, $\left<p_{T}\right>$). The fused string conserves energy-momentum and color charge. Two immediate implications of string fusion are a reduction in overall particle multiplicity and an increase in $\left<p_{T}\right>$ of produced hadrons. Both these effects are observed in Au+Au data at RHIC \cite{bib7}. Hadronization of the strings proceeds with a Schwinger probability dependence \cite{bib5}. There is no a priori assumption of a phase transition in either the DPM or PSM. However, the clustering of strings may eventually lead to the formation of a macroscopic cluster at a certain critical density. This would indicate the onset of the percolation phase transition \cite{bib8,bib9}. In this context, string fusion can be interpreted as an intermediate stage leading toward the formation of deconfined quark-gluon matter \cite{bib10}.

A linear relationship has been found in high-energy colliding hadron experiments between the multiplicity in a forward $\eta$ region ($N_{f}$) and average multiplicity in a backward $\eta$ region ($N_{b}$)\cite{bib15}:

\begin{equation}\label{linear} 
<N_{b}(N_{f})> = a + bN_{f} 
\end{equation}
The coefficient b is referred to as the correlation coefficient and is a function of incident energy and atomic number. It can be expressed in terms of the expectation value \cite{bib4}:

\begin{equation}\label{b}
b = \frac{<N_{f}N_{b}>-<N_{f}><N_{b}>}{<N_{f}^{2}>-<N_{f}>^{2}} = \frac{D_{bf}^{2}}{D_{ff}^{2}}
\end{equation}
where $D_{bf}^{2}$ and $D_{ff}^{2}$ are the backward-forward and forward-forward dispersions, respectively. This result is exact and model independent \cite{bib12}.

Short and long-range (in rapidity) multiplicity correlations are predicted as a signature of string fusion \cite{bib10,bib11}. When strings fuse, a reduction in the long-range forward-backward correlation is expected. The existence of long-range multiplicity correlations may indicate the presence of multiple partonic inelastic collisions. These correlations arise from the superposition of a fluctuating number of strings, such that \cite{bib12}:
\begin{equation}\label{dbf}
D_{bf}^{2} = <N_{f}N_{b}>-<N_{f}><N_{b}> \propto \left[\left(<n^{2}>-<n>^{2}\right)\right]<N_{0f}><N_{0b}>
\end{equation}
with $(\left<n^{2}\right>-\left<n\right>^{2})$ the fluctuation in the number of inelastic collisions and $\left<N_{0f}\right>, \left<N_{0b}\right>$ are the average multiplicity produced from a single inelastic collision.
Therefore, $D_{bf}^{2}$ should be sensitive to the presence of long-range multiplicity correlations.

In this paper we discuss the results from 200 GeV Au+Au collisions for all charged particles with $p_{T}$ in the range from 0.1 to 1.2 GeV/c, to ensure a sampling of soft particles only. To eliminate short-range correlations, a gap in pseudorapidity ($\eta$) of 1.6 units is considered \cite{bib4}. The forward pseudorapidity interval was $0.8 < \eta < 1.0$ and the backward was $-1.0 < \eta < -0.8$. 

\section{Data Analysis}\label{data}  

The data utilized for this analysis is from year 2001 (Run II) $\sqrt{s_{NN}}$ = 200 GeV Au+Au collisions at the Relativistic Heavy Ion Collider (RHIC), as measured by the STAR (Solenoidal Tracker at RHIC) experiment \cite{bib13}. The main tracking detector at STAR is the Time Projection Chamber (TPC) \cite{bib14}. The TPC is located inside a solenoidal magnet generating a constant, longitudinal magnetic field. For this analysis, data was acquired at the maximum field strength of 0.5 T. All charged particles in the TPC pseudorapidity range $0.8 < |\eta| < 1.0$ and in the $p_{T}$ range 0.1-1.2 GeV/c were considered. The $\eta$ range was chosen to eliminate the effects of short-range correlations from sources such as cluster formation, jets, or resonance decay. The collision events were part of the minimum bias dataset. The minimum bias collision centrality was determined by an offline cut on the TPC charged particle multiplicity within the range $0.5 < |\eta| < 0.5$. The centralities used in this analysis account for 0-10, 10-20, 20-30, 30-40, 40-50, 50-60, 60-70, and 70-80\% of the total hadronic cross section. An additional offline cut on the longitudinal position of the collision vertex ($v_{z}$) restricted it to within $\pm 30$ cm from $z=0$ (center of the TPC). A combined two million minimum bias events satisfied these requirements and were used for this analysis. Corrections for detector geometric acceptance and tracking efficiency were carried out using a Monte Carlo event generator and propagating the simulated particles through a GEANT representation of the STAR detector geometry. 

In order to eliminate the effect of statistical impact parameter (centrality) fluctuations on this measurement, each relevant quantity ($\left<N_{f}\right>$, $\left<N_{b}\right>$, $\left<N_{f}\right>^{2}$, and $\left<N_{f}N_{b}\right>$) was obtained on an event-by-event basis as a function of STAR reference multiplicity, $N_{ch}$. A linear fit to $\left<N_{f}\right>$ and $\left<N_{b}\right>$, or a second order polynomial fit to $\left<N_{f}\right>^{2}$ and $\left<N_{f}N_{b}\right>$ over the range $0 \leq N_{ch} \leq 600$ was used to extract these quantities as functions of $N_{ch}$. Tracking efficiency and acceptance corrections were applied to each event. These were then used to calculate the backward-forward and forward-forward dispersions, $D_{bf}^{2}$ and $D_{ff}^{2}$, binned according to the STAR centrality definitions and normalized by the total number of events in each bin.
 
\begin{figure}[t]
\centering
\vspace*{1cm}
\epsfysize=3.5in
                 \epsffile{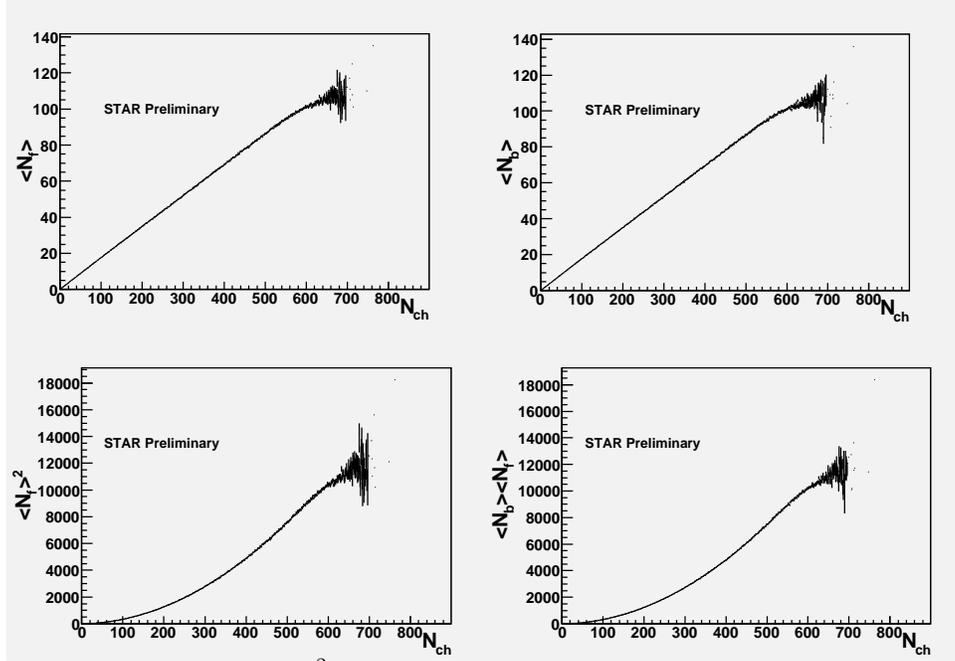}
\vspace*{-0.5cm}
\caption[]{$\left<N_{f}\right>, \left<N_{b}\right>$, $\left<N_{f}\right>^{2}$, and $\left<N_{f}N_{b}\right>$ as functions of STAR charged particle reference multiplicity ($N_{ch}$). The quantities are found for event-by-event for all values of $N_{ch}$.}
\label{fig1}
\end{figure}

\section{Results}\label{results}

Fig. \ref{fig1} presents the mean multiplicity $\left<N_{f}\right>$, $\left<N_{b}\right>$, in the forward and backward pseudorapidity region ($0.8 < |\eta| < 1.0$) along with $\left<N_{f}\right>^{2}$ and $\left<N_{f}N_{b}\right>$ as functions of reference multiplicity, $N_{ch}$. The results are calculated event-by-event for every unit $N_{ch}$. The $\left<N_{f}\right>$ and $\left<N_{b}\right>$ demonstrate a linear dependence over much of the centrality range. Only statistical uncertainties are present and are less than 1\%, except for the highest multiplicities. These results are uncorrected for tracking efficiency and acceptance. Due to statistical limitations, it is not possible to apply corrections for every value of $N_{ch}$. One value for the correction, calculated for each centrality, is applied to the resultant values $\left<N_{f}\right>, \left<N_{b}\right>, \left<N_{f}\right>^{2}$, and $\left<N_{f}N_{b}\right>$ for every event in that centrality. Therefore, all events falling within a particular centrality have the same correction.

\begin{figure}[t]
\centering
\vspace*{1.0cm}
\epsfysize=4.3in
                 \epsffile{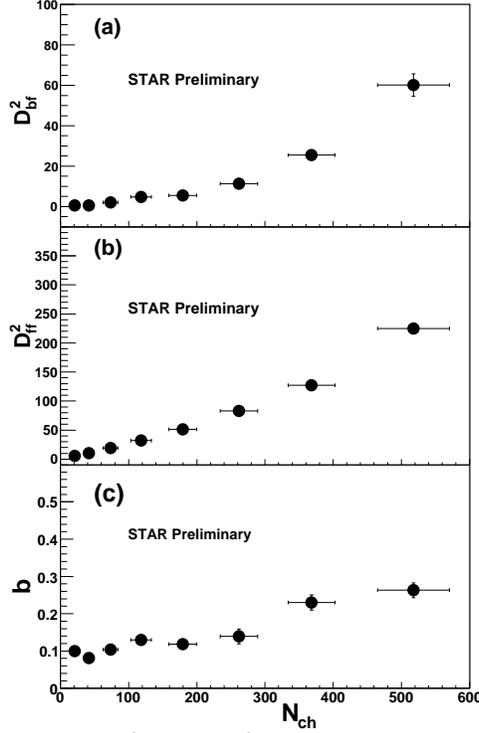}
\vspace*{-1.5cm}
\caption[]{(a) $D_{bf}^{2}$, (b) $D_{ff}^{2}$, and (c) b from Au+Au collisions at $\sqrt{s_{NN}}$ = 200 GeV, as a function of the STAR reference multiplicity, $N_{ch}$. Vertical error bars represent statistical and systematic uncertainties, horizontal, the RMS of the mean multiplicity distribution in each centrality bin.}
\label{fig2}
\end{figure}

Fig. \ref{fig2} shows the results for $D_{bf}^{2}$, $D_{ff}^{2}$, and the correlation coefficient, b, as a function of $N_{ch}$ for the 8 centrality bins. The presence of long-range multiplicity correlations are evident from $D_{bf}^{2}$ and b. The growth of $D_{bf}^{2}$ as a function of $N_{ch}$ is consistent with an increasing long-range correlation from peripheral to central heavy-ion collisions, corresponding to a greater number of fluctuating strings. The linear evolution of $D_{ff}^{2}$ as a function of $N_{ch}$ represents a (short-range) auto-correlation of particles within the same window. The b coefficient represents the long-range correlation normalized to the short-range correlation in the forward $\eta$ window. Were there no long-range correlations ($D_{bf}^{2} = 0$), then b would vanish. The vertical error bars in Fig. \ref{fig2} are statistical and systematic uncertainties. The statistical errors are smaller than the data points. The systematic errors are determined by varying cuts on $|v_{z}|$, number of fit points on track in the TPC, fit ranges for $\left<N_{f}\right>, \left<N_{b}\right>, \left<N_{f}\right>^{2}$, and $\left<N_{f}N_{b}\right>$, etc. Horizontal error bars are the RMS value from the mean multiplicity for each centrality.

\begin{figure}[t]
\centering
\vspace*{-.3cm}
\epsfysize=4in
                 \epsffile{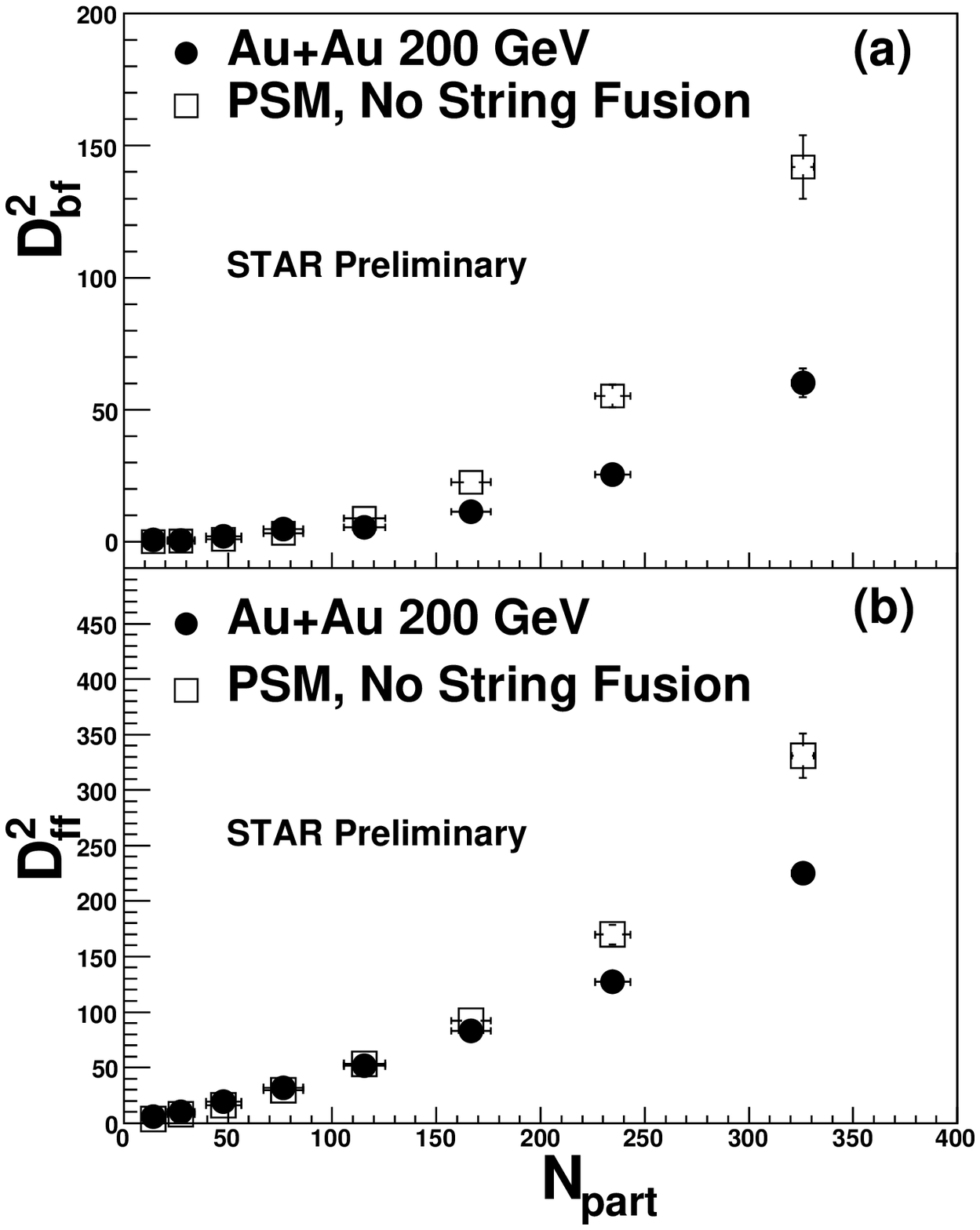}
\vspace*{-1.5cm}
\caption[]{(a) $D_{bf}^{2}$ and (b) $D_{ff}^{2}$ as a function of the $N_{part}$ in 200 GeV Au+Au collisions, compared to the PSM with no string fusion (independent strings). There is good agreement for both quantities in peripheral collisions. A large discrepancy exists in central collisions, suggesting an underlying dynamical mechanism that reduces $D_{bf}^{2}$ and $D_{ff}^{2}$.}
\label{fig3}
\end{figure}

\begin{figure}[t]
\centering
\vspace*{-.3cm}
\epsfysize=4in
                 \epsffile{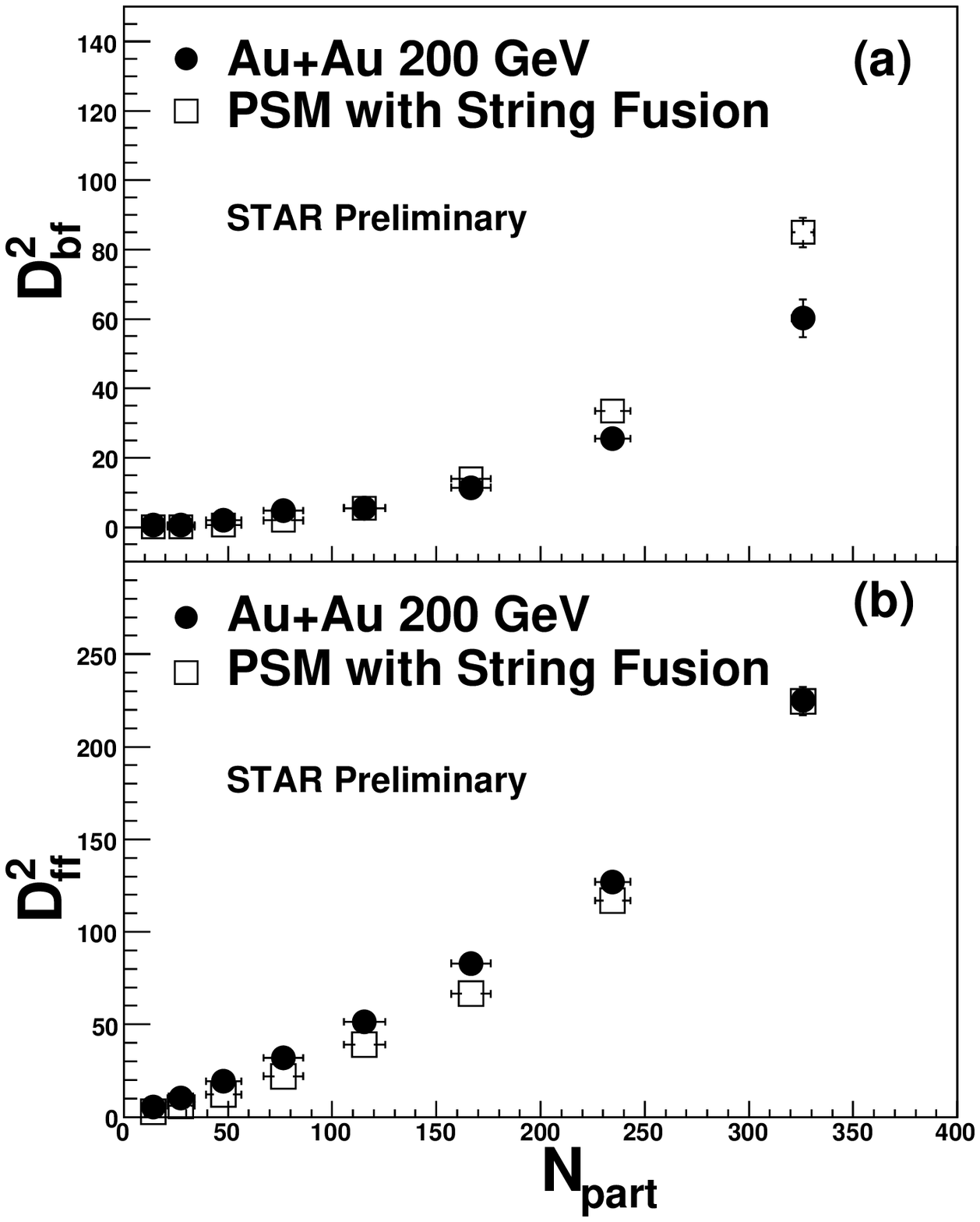}
\vspace*{-1.5cm}
\caption[]{(a) $D_{bf}^{2}$ and (b) $D_{ff}^{2}$ as a function of the $N_{part}$ in 200 GeV Au+Au collisions, compared to the PSM with string fusion (collective strings). There is good agreement for both quantities in peripheral collisions. The discrepancy in central collisions is reduced compared to Fig. \ref{fig3}, but still persists. This remaining difference hints that an additional reduction, beyond that offered by two soft string fusion, is required.}
\label{fig4}
\end{figure}

The comparison of $D_{bf}^{2}$ and $D_{ff}^{2}$ as calculated from the Au+Au data to that from the Parton String Model (PSM) with the string fusion on or off is presented in Figs. \ref{fig3} and \ref{fig4}, respectively. The PSM (with two string fusion) minbias multiplicity distribution closely matches that of the corrected Au+Au data. It also describes the $\left<p_{T}\right>$ enhancement, particle ratios, strangeness production, etc., seen in Au+Au collisions at RHIC \cite{bib17}. As such, the centrality cuts in the PSM correspond to the same percentage of the minimum bias cross section as in the data and are plotted as a function of the number of participant nucleons in the collision ($N_{part}$), calculated from Monte Carlo Glauber model \cite{bib16}. The statistical and systematic uncertainties in Figs. \ref{fig3} and \ref{fig4} are the same as those described for Fig. \ref{fig2}. 

Figs. \ref{fig3} and \ref{fig4} show good agreement in peripheral collisions between data and the independent (no string fusion) or collective (with string fusion) model. In central Au+Au collisions, $D_{ff}^{2}$ with the independent string description deviates from the data, but shows excellent agreement for fusion of two soft strings. This confirms the agreement in multiplicity between the PSM (with two string fusion) and data, as $D_{ff}^{2}$ is a multiplicity effect. However, there is a large discrepancy in $D_{bf}^{2}$ for central collisions for both the independent and collective PSM, compared to the data. This discrepancy is greater for the case of independent strings (Fig. \ref{fig3}). This suggests an additional, dynamical reduction in the number of particle sources in central Au+Au collisions, greater than that provided by the fusion of two soft strings (Fig. \ref{fig4}).

\section{Conclusions}\label{concl}
 
We have presented data on long-range multiplicity correlations of soft particles in 200 GeV Au+Au collisions. A long-range multiplicity correlation is seen for all centralities and increases from peripheral to central collisions. Comparison to the Parton String Model (PSM) using both interacting and non-interacting strings shows good agreement for peripheral collisions, but large differences in central collisions. The reduction in the long-range multiplicity correlation, $D_{bf}^{2}$, even compared to the collective PSM (fusion of two soft strings), is a possible indication of an additional, collective effect in central Au+Au collisions at 200 GeV.

\bibliography{lajolla2006-template}
\bibliographystyle{lajolla2006}
 
\vfill\eject
\end{document}